\documentclass[aps,prd,showpacs,nofootinbib,preprint,tightenlines]{revtex4}

\def\be{\nopagebreak[3]\begin{equation}}
\def\ee{\end{equation}}
\def\ba{\nopagebreak[3]\begin{eqnarray}}
\def\ea{\end{eqnarray}}

\def\l{\langle}
\def\r{\rangle}

\newcommand{\teta}{\rlap{\lower2ex\hbox{$\,\tilde{}$}}\eta{}}

\newcommand{\y}{\hat{y}}              

\newcommand{\py}{{\hat{\pi}^{(y)}}}     
\newcommand{\pyRI}{\hat{\pi}^{(y)R,I}}

\newcommand{\pf}{\hat{\pi}}           
\newcommand{\fluc}[1]{(\Delta #1)^2}      
 
\newcommand{\cre}{\hat{a}^{\dagger}}    

\newcommand{\ann}{\hat{a}}            
\def\be{\nopagebreak[3]\begin{equation}}
\def\ee{\end{equation}}
\def\ba{\nopagebreak[3]\begin{eqnarray}}
\def\ea{\end{eqnarray}}

\def\l{\langle}
\def\r{\rangle}

\begin{document}

\title{The seeds of cosmic structures as a door to Quantum Gravity Phenomena}

\author{ 
 Daniel Sudarsky\\ \small
 {\it Departamento de Gravitaci\'on y 
Teor\'\i a de Campos, Instituto de Ciencias Nucleares} \\ \small
{\it Universidad Nacional Aut\'onoma de M\'exico} \\ \small {\it 
Apdo. Postal 70-543 M\'exico 04510 D.F, M\'exico} }

\begin{abstract} 
This paper  contains a critique of the standard inflationary account of the origin of cosmological 
structures from quantum fluctuations in the early universe.  This critique can be thought
 to be purely 
philosophical in  nature, but  I  prefer to view it, rather, as  arising from the need to put the interpretational
aspects of the theory -which quite obviously  lie at the basis of any comparison with experiments- on  the firm grounds required by the  unique features of the problem at hand.
 This discussion is followed by a proposal to complement that treatment to deal  with the unsatisfactory aspects of the standard account of the problem,   using  Penrose's ideas about the quantum gravity induced collapse of  the quantum states  of matter fields.  The formalism developed to carry out this analysis was first introduced in  [1]
and leads to unexpected predictions and to novel avenues
 to confront  some of the  details of the proposal with observations.  In my view, this is, therefore, the most promising path towards quantum gravity phenomenology.

 \end{abstract}

\maketitle

\section{The Problem}

The inflationary theory of the origin of structure in our universe, although generally acclaimed for its successes \cite{Guth, CMB}, has a very 
unsettling aspect:  it does not account for the transition from a homogeneous and isotropic (H.\&I.) early state of the universe to its late, 
anisotropic and in-homogeneous state.
  What do we mean by that? Isn't this, precisely, what the account we read in books and articles  does in calculating the spectrum of primordial fluctuations?  We will see that the answer one gives to this question  depends very strongly on what one expects a physical theory to be able to deliver,  an issue which is strongly correlated with one's views about what  quantum mechanics is and what  it is not.   The noteworthy point is, however, that  despite what it might seem at first sight,  the way these issues  are approached have implications that transcend the purely philosophical aspects and impinge on 
   our understanding of one of the fundamental questions in modern cosmology.   
  But let us go back to the questions raised:
  does the standard inflationary scenario, and the accompanying calculations of the  spectrum of primordial fluctuations, really account for the transition from a H.\&I. early state of the universe to its late, anisotropic and inhomogeneous state\footnote{ This point is sometimes characterized as the ``transition from the quantum regime to the classical regime", but I find this a bit misleading: most people would agree that  there are  no classical  or quantum regimes. The fundamental description ought to be quantum  always, but there are regimes in which certain quantities can be  described  to a  sufficient accuracy by their classical counterparts representing  the  corresponding expectation values.  This depends of course on the physical state, the underlying dynamics, the quantity of interest, and the context in which we might want to use it.}?  If the answer  to the above question depends on one's view of quantum theory, why does this article start with  such a categorical assertion,  that the answer is in the negative? My point will be that the views that one would be led  to take  about quantum mechanics, in seeking a justification to answer in the positive,  are not  one that  physicists would like to take, are not one that would justify its use in the situation at hand, and/or are not one that
 would allow one  to consider the sort of issues raised in the search for quantum gravity.
  
  This article will be devoted, to a large extent, to deal with the issues above, but will include a brief description of the richness in
    perspective that  can be achieved if one takes a stricter ontological view of
  the essence of quantum physics. We will see that not only will we gain the right to ask more profound questions, but that, in 
  attempting to deal with them, we will be led to new and  surprising  insights about the nature of quantum gravity and their relevance to
   the understanding of the early universe.  The fact that issues that can  initially be deemed to be  merely philosophical, can transcend
    such domain and enter  the 
  realm  of observational  scientific inquire, can  be expected to come as a  shock to  people  belonging to the generations of physicists trained not to
   ask that type of questions. However, these questions are indeed  some of the hardest ones,  and  it is the hope of this author  that this manuscript may  contribute to  the changing of  the above mentioned  unfortunate trend.

\section{What is quantum physics?}

It must be  surely  pretentious beyond belief to raise this issue in the company of prestigious physicists
  that have devoted their lives to the even more complex problem of making gravitation and quantum physics compatible. 
   These efforts have been concentrated in the highly nontrivial  development of the mathematical language and the
   subtleties that arise in constructing plausible theoretical models applicable to quantum gravity. Perhaps, it is the enormity of the task and the
    difficulty of that challenge, that having absorbed the attention of the most brilliant minds in our field, have left as relatively much less explored
     the issues which I will be rising.    
     
     These are related to the so called  ``measurement problem in quantum mechanics" a subject that has 
     puzzled physicists and philosophers of physics  from the time of the inception of the theory 
     \cite{MPQM}.  We will however focus  here on one particular  instance of   of the problem:  the cosmological setting, which is  a subject that has received  much less attention from the physics community.   There are of course notable exceptions to the assessment above, represented by
    thinkers like R. Penrose, \cite{Penrose1} J. Hartle \cite{Hartle1},  and others who have faced 
    these and related issues with  very open and  critical minds. Having said  all this, I must nevertheless try to answer the question posed, 
    if for no other reason,  just  to  make my position clear to the reader, and help him understand the hows and whys of the postures taken in this manuscript.
     To do this I will go through some of the postures that I have encountered on the subject:
    
     a)  {\it Quantum physics  as  a complicated theory of  statistical physics.}
      By this, I am referring to the position that holds that quantum mechanics acquires meaning only as it is applied to  an ensemble of  identical systems. 
       In this view, one must accept that a single atom in isolation is not described by quantum mechanics. Let us not get confused by the correct but simply 
       distracting argument that atoms in isolation do not exist.  The point is whether,
       to the extent to which we do  neglect its interactions with distant atoms,  and specially  with the electromagnetic field which, even in its vacuum state is known to interact  with  the atom, quantum mechanics is applicable to the description of a single atom. Again, what can we mean by that, if we know that in order to be able to say anything about the atom, we must   make it interact with a  measuring device?  Well, the question is simply whether applying the formalism of quantum mechanics  to treat the isolated atom can be   expected to yield correct results  as it pertains the subsequent measurements.?  One might  think that this is a nonsensical question, as these results are always statistical in nature. The point is that this  statement is not really accurate: for instance if the atom (say, of  hydrogen) was known to  having been prepared  in its ground state, the probability of measuring any energy other than the one in the ground state is zero. 
   Furthermore,  the probability of observing
        the atom in a high  angular momentum   eigenstate is  equally zero. In fact, for any observable commuting with the hamiltonian  the predictions are not statistical 
        at all, but 100\% deterministic and precise! If so, there must be something to the description of that single atom by its usual  quantum mechanical state, 
       and thus it becomes blatantly false the notion that quantum mechanics can not be applied to single system. What is true, of course,  is that, in applying the theory to a single system, the  predictions we  can make with certainty are very  limited,  with the extent of such limitations being determined both by the nature of the system's 
          dynamics, and by the way the system was initially prepared. Moreover, in relation with the issue that concerns us in this article, taking  a posture
           like this, about quantum physics,  would be admitting  from the beginning that we have no right to employ such theory in addressing questions concerning the  unique   
            universe to which we have access, even if we were to accept that somehow there exists an ensemble of universes  to which we have no access. Note, moreover, that  we should beware from confusing statistics of universes and statistics within one universe.  Furthermore, if a quantum state serves only to represent an ensemble,  how is  each element of the ensemble to be described?  Perhaps, it can not  be described at all ? What do we do in that case with our universe?
      
      b) {\it Quantum physics as  a theory of human knowledge}. Within 
     this  view of quantum theory,  the state of a quantum system does not reflect something about the system, but just what we know about the system.  Such view, naturally rises the question what is there to be known about the system if not something that pertains to the system?  The answer comes in the form  of: correlations between the system and  the measuring devices, but then,  what is the meaning of these correlations?. The usual meaning of  the word correlation implies some sort  of  coincidence of certain conditions pertaining  to one object with some other conditions pertaining to the second object.    However, if a quantum state describes such correlation,  there must be  some meaning  to the conditions pertaining to  each  one  of the objects. 
         Are not these, then,  the aspects that are described by the quantum mechanical state for the object? If we answer in the negative,  it must mean that there  are further descriptions of the object that can not be casted in the quantum mechanical state vector. On the other hand,  if we answer  in the positive we are again taking 
          a view whereby the state vector says something about the object in itself. Perhaps  we are just going in circles. For those  who read these  considerations as 
           philosophical  nonsense, let us just say that if we follow this view, we have no right to consider  questions about the evolution of the universe in the absence of sapient beings, and much less to consider the emergence, in that universe,  of the conditions that are necessary for the eventual evolution of humans, while using a quantum theory. I would take it even further: what would be  the justification for considering  states in any model of  quantum gravity, if we took such view of quantum physics?

      c) {\it Quantum physics  as  an non-completable  description of the world.}  With this I am  referring to any posture that effectively, if not explicitly,  states:``The theory is incomplete, and no complete theory containing it exists or will ever do" . This view  will  be  considered as  being held  by  the many physicists that, while not openly advocating such posture, will direct us to use quantum mechanics " as we all know how "  while reminding us with  a stern voice that no violation of quantum mechanics has ever been observed.   While this  is with not  doubt  a  literally correct statement, we must remind our colleagues that by this,  
they refer, of course,  to the rules as found in any quantum mechanics  text book, that  essentially rely on   to the Copenhagen interpretation, which as we all know,  raises  severe  interpretational issues  that  {\bf become insurmountable  once we leave the laboratory and consider applying quantum theory to something  like the universe itself}.  According to this view, we should content ourselves using its tools, and making, in the situation at hand,  non-rigorous predictions\footnote{ Quantum theory gives perfectly well defined and rigorous predictions,  which are in general of a statistical nature,  once we have identified the state of the system the observable that is going to be measured, which must include in general  the time at which  the measurement occurs.  The  in the standard treatments  of  the situation at hand no such  explicit  identifications are  made  and thus the line of reasoning connecting the formalism to the predictions is best described as non-rigorous.  }.
%
  We must acknowledge, however, that in  situations where one  can not point to the classical-quantum dividing line, where we can not identify the system and the apparatus, nor the observables that are to be measured, the entity carrying out those  measurements and the time  at which the measurements are to be thought as taking place, we have,  in fact, no clearly defined scheme specifying  how to make the desired predictions. That is, in dealing with the  questions pertaining to the early universe in terms of quantum theory, we  have no clear and specific rules for making predictions.  However, according to the colleagues arguing for such  practical posture, we should be content  with the fact that the predictions have in fact been made, and that they do seem to agree with observations. The issue is, of course, that in the absence of a well defined set of rules, that are explicit to the point where a computer could,  in principle, arrive to the predictions using only the explicit algorithm and the explicitly stated inputs, we have no way to ascertain whether or not,  such "predictions" do  or do not,   follow  from the theory. We can not be sure whether or not some unjustified  choices, manipulations  and arguments have   been used as part of  the process by which the predictions have been obtained. Correct predictions are not  enough\footnote{ For those that shiver upon the last statement let me recall a theory about venereal diseases that was popular in ancient Rome:  venereal diseases come from Venus, the Goddess of Love, obviously. The celestial enemy of Venus was Mercury,  which  constantly challenged the Goddess's  for the attention of the Sun (Apollo).  Fortunately for humans, Mercury had  a substance associated with him, that could be found on Earth:  the metal mercury. It was  then evident that substances containing mercury should help in fighting the  diseases associated with Venus.  And it was a known fact ( at least among the practitioners of medicine) that certain  mercury salts  
      did provide substantial relief for those infected with the most common of the venereal diseases of the time. I do not, of course, pretend to compare any theory of modern physics with ancient Rome's medical arguments. My point is only to illustrate the statement I made above:  correct predictions are not enough.} ! They must first be  actual predictions.  This is quite clear, particularly, when the argumentative connections used in arriving to the predictions  are so loose that no one can be sure whether these are in fact predictions or not.  Specially suspect are,  of course, those "predictions" that are, in fact, retrodictions, and on this point we should be aware that long before inflation  was invented,  Harrison and Zel'dovich \cite{HZ} had already concluded the form of the primordial spectrum, based  on rather broad observations about the nature of the large scale structure of our universe. 
      
      d) {\it Quantum physics as  part of a  more complete  description of the world.}  Here we are not referring to an  extension of the theory  into some sort of hidden variable type,  as the problem we want to rid the theory from is not its indeterminism, but the so called measurement problem. 
      Completing  the theory would require something  that removes the need for a  external measurement apparatus, an external observer, etc.  There  are, for instance,  ideas  like  generalization of quantum physics using a scheme  based on  sums over decoherent  histories  proposed by J. Hartle  \cite{Hartle1}, others invoking something like the  dynamical reduction models proposed   by Ghirardi, Rimini \& Weber \cite{GRW}, and 
      the ideas of   R. Penrose about the role of gravitation in modifying quantum mechanics in the
      merging of the two aspects of physical reality \cite{Penrose1}.  The present manuscript, as well as  the original treatment of these issues carried out in collaboration with A. Perez and H. Shalmann \cite{Us},are inspired on  Penrose's generic ideas.
      
      e) {\it Quantum physics as  a complete  description of the world.}   The view that quantum mechanics faces no open issues and that, in particular, the measurement problem  has been solved.
          Among the holders of these views one can further identify two main currents:  those that  subscribe  to  ideas along the so called  "many world interpretation of quantum  mechanics" and consider this to be a solution to the measurement problem, and  those that hold a view  that the measurement problem in quantum mechanics has been solved by  the consideration of    "decoherence".  Let us first note that the many world interpretation does very little to ameliorate the measurement problem, as there is a mapping between what in that approach would be called  the splittings of worlds, and what  would be  call  "measurements" in the Copenhagen interpretation. Thus every question that can be made in the latter interpretation  has a corresponding one in the many worlds interpretation.  For the case of the measurement problem the issues would be.  when does a  world splitting occur? why, and under what circumstances does it occur?   What constitutes a  trigger? Concerning the decoherence mechanism as  a solution to the measurement problem I would like to start  by quoting the postures that in these regards are held by  several people that have considered the issue at length in order to dispel the  widespread notion that such is the consensus view:
    
    {\it  "Many physicist nowadays think that decoherence provides a fully satisfying answer to the measurement problem. But this is an illusion."} Arnold Neumaier \cite{Neumaier}.
      
    {\it " ...note that the formal identification of the reduced density matrix with a mixed state density matrix is easily misinterpreted as implying that the state of the system can be viewed as mixed too..
   ..  the total composite system is still described by a superposition, it follows from the rules of quantum mechanics that no individual  definite state can be attributed to one of (the parts) of the system ..."}, M. Schlosshauer \cite{Schlosshauer}.
      
      
    {\it  "Does decoherence solve the measurement problem? Clearly not.  What decoherence tells us is that certain objects appear classical when observed, But what is an observation? At some stage we still have to apply the usual probability rules of Quantum Theory."}  E. Joos  \cite{Joos}.

   In fact, we will see that when dealing with cosmology the problem becomes even more vexing and acute.   Nonetheless, most researchers in the field seem to take some version of decoherence as  the paradigm where the  direct application of  the standard forms of quantum mechanics to the problem  at hand finds its justification. Significantly, the diversity of precise approaches 
   indicates some degree of in-satisfaction  of some researchers with the views 
   of others \cite{Cosmologists}.    
   
 Before engaging on the cosmological case,  let us review briefly what decoherence is, and what can and can not do.  
 
    Decoherence is the process by which a system that is not isolated, but in interaction with an environment (as are all physical systems except the universe itself) "looses" or " transfers" coherence into the degrees of  freedom of such environment. It is a well studied effect that follows  rather than supersedes  the laws of quantum physics.  It is, therefore, clear that, in principle, it can not be thought to offer  explanations  that  go beyond what  can be directly inferred from the application 
    of the principles of quantum physics. Its main achievement is to allow for the studying the conditions in which the quantum interferences expected 
    from the idealized  consideration of a system as isolated, become observationally suppressed as the result of the system's  interaction with the environment. 
    
The basic  recipe for an analysis of decoherence in a given situation follows the following steps: 
    
 1) Divide D.O.F. : system + environment
  ( identify inaccessible or irrelevant D.O.F.).

2) Compute  Reduced Density matrix
    (trace over environment D.O.F.).

3) Perform suitable time average so that the off-diagonal matrix elements vanish.

4) Regard the diagonal density matrix as describing a statistical ensemble.

    {\bf The Problems:}  once one has understood  why  certain interferences can not be observed  in practice, it is tempting to conclude that one has understood the "emergence of classicality", and 
    that therefore there is nothing  left of the so called  " measurement problem" in quantum mechanics. This  turns out to be  a simplistic  and misguided  conclusion, as indicated by the quotations listed  above. There are  in fact, at least two  very serious problems  with considering decoherence  in this light: 
    
    I) The  basis problem: it is clear that the diagonal nature of the  reduced density matrix obtained in the step 3) of the  program above,  will be lost, in general, upon a change of basis for the Hilbert space of the system at hand.  This is taken  to mean that the nature of the  system-environment interaction selects a so called pointer basis which underscores the aspects  that have become classical as  a result of decoherence. The point of course is that this leaves one with the usual situation whereby, if the selected basis is, say,  the position basis, the momentum of the system remains undetermined and thus one
    can not argue that classicality has really emerged.  
    
            II)The definite outcomes problem:  here the problem is  the absence of sufficient justification for the interpretation of the mixed state described by the density matrix as describing   a statistical ensemble and in regarding  a single system as being in  definite   yet unknown state among the ones represented in the diagonal elements of the density matrix.  The result  that emerges  from the decoherence  calculations rather   indicates that the system must be regarded as coexisting in the various alternatives, but where the interferences in  the  observables associated with the particular  pointer basis being suppressed.  Selecting among these   alternatives  can be viewed  as deciding  between  the " choice vs coexistence" interpretations. In order to argue that decoherence really leads to the emergence of classicality one would have to advocate  the "choice" interpretation, but  well known, and experimentally confirmed   aspects of quantum mechanics  such as  the violation of Bell inequalities, force us to opt for the "coexistence"   interpretation \cite{Bell, Aspect}.

           The next example,  from ordinary non-relativistic  quantum mechanics,   serves as a clear analogy of the situation we face:
            consider a single particle in a state corresponding to a  minimal  wave  packet centered 
            at $ \vec X = ( D, 0, 0)$ ( the vectors in 3-D space are given in cartesian coordinates  (x, y, z)).
             Let the particle have its spin pointing in the $+y$ direction.  Take this state and  rotate it by and angle $\pi$ about the $z$ axis.  Now consider the superposition of the initial and the rotated states. The resulting state   is clearly  symmetric under rotations by $\pi$ around $z$.
         Now consider taking the trace  over the spin D.O.F.
The resulting density matrix is diagonal.
 Can we say that  the situation  has become classical? Of course not.   Is the state still invariant under rotations of 
magnitude $\pi$ about the $z$ axis? Obviously yes.
Can a mathematical manipulation with no physical 
process counterpart ever change the state of the system? Answering yes would take you to the view discussed in b).

\section{ The exacerbated problem: applying quantum physics to the early universe}

  We should point out that some   researchers in the field, such as \cite{Padmanaban},  have acknowledged that there is something  mysterious in the standard account  of the  emergence of structure, 
  and people  like J. Hartle \cite{Hartle1}  that have pointed out the need to generalize quantum mechanics to deal  with cosmology, and of course  R. Penrose, who in his last book  \cite{RoadToReality} has stressed  the relevance of the general measurement problem in quantum mechanics to the problem of breakdown of the H.\&I.  during inflation,  comparing  it with the  problem of the breakdown of spherical symmetry in a particle decay.
In my view, this  analogy  does not emphasize the point that, in the cosmological context,  the problem is  even more severe than in ordinary situations, because, in that  case,  we can not even rely on the  strict  Copenhagen interpretation  as a source of  safe practical rules. 

  As an example that exhibits quite  clearly  the deepening of the problem in this context, let's consider the following  quotation from  a  well known thinker on these sort of   issues in  quantum theory:
   {\it  "As  long as we remain within the realm of mere predictions concerning what we  shall observe (i.e. what will appear to us) and refrain from stating anything concerning ``things as they must be before we observe them " no break in the linearity of quantum dynamics is necessary. "}  D'Espagnat \cite{DEspagnat}.
    
{\it \bf However, in the cosmological setting, we need  to deal precisely with this situation:  we need to think about the state of affairs of the universe before the emergence of the conditions that make us possible, before we existed and before we ever carried out an observation or measurement. }

In the cosmological setting,
we seek a historical, that is a time  development,  description of cosmic evolution that follows the laws of physics (would this  be, perhaps, too much to ask of a physical theory of cosmology?).
Such description  should explain how did WE arise, in a path covering the emergence of the primordial density fluctuations, of galaxies, stars, and  planets,  and eventually living organisms, humans, cultures, etc..
Such an account should not rely on the measurement  (in) abilities of the late evolved creatures to explain the  emergence of  conditions that make them possible.  From this perspective,  one can not justify identifying some D.O.F as irrelevant environment, based on the current, or even permanent,  limitations of humans, in  the analysis of the emergence of the primordial  density fluctuations, for doing so  leads to a circular argument with no explanatory value.

 Furthermore, one might  be asked,  when attempting to  follow some of the standard versions of the explanations of these delicate points,  to accept one or more  of the following notions:
1)that quantum physics does not describe our universe, as it was never H.\&I.,  only   a certain  ensemble  of universes was. The argument here seems to rely on the notion of such ensemble of universes could be used to describe
 aspects of our universe that could not be described in isolation. The line of thought would indicate that 
 it is the superposition of the states of all  the universes in the ensemble, what is 
 represented by the H.\&I. quantum state.  The problem is that this is not quantum mechanics:  if we have two systems, each of which is described by a quantum state, the composite  system is not described by the superposition of the two quantum states ( it is described by their direct or "tensor"  product in the direct product of the corresponding Hilbert spaces). 
2) that our universe is still H.\&I., and that the appearances to the contrary are the result of our inability to observe all the degrees of freedom. This posture makes us and our limitations,  a fundamental part of the explanation  of the origin of the conditions that  makes our existence possible. 3)
That "it does not matter when the universe stopped being H.\&I. ", without being able to even address  issues  such as {\it when?, why?}, or  {\it due to what?}.

I find it quite remarkable that many physicists  seem to look for, or content themselves with, what I would call technical  pseudo-answers to the problem. These approaches can be  globally characterized as calculation that while technically correct  fail to address the issue at hand by the implicit acceptance of an interpretational scheme of the formalism, which is not  justified in the present context. 
 These often include pseudo-analogies, that is, comparisons of the present problem with other problems  which superficially seem complete  analogies  but that upon careful examination  reveal that they are missing crucial aspects, particularly those that make the problem at hand such a challenge.
 The danger of the analysis' 
 by analogy is that they often fail to dissect the problem to the point where all aspects of the analogy  have been explicitly exposed and  can therefore be examined.   Consider, for instance,  a recent  article \cite{Jerome 1} which  proposes  an analogy between the  process that lies at the origin of  the anisotropies and inhomogeneities in the early universe, with the process of particle creation out of the vacuum in the presence of a sufficiently intense electric field,  a process known as   the Schwinger process \cite{Schwinger}.  In that calculation, one evaluates the  $S$ matrix element between the "in" vacuum and the "out" vacuum,  and interprets the difference between  the result and the unity as a measure of particle creation. This identification would be  justified  by unitarity and the observation that the $S$ matrix element between the "in" vacuum and the other vectors in the "out" Hilbert space, such as a specific state with  a single  pair of electron   and positron, would  be  interpreted as the
 probability of  pair creation in such state. This  would in turn,   be justified  by the standard quantum mechanics postulate of the projection postulate, indicating that $ | \l A  | B \r |^2$ as the probability  of finding the system in the state $|A \r $, if it was originally prepared in the state $| B \r$.  But this assumes  that an observation,  or measurement is carried out for an observable  for which $|A \r $   is  a  (non-degenerate)  eigenstate.  In the absence of a measurement, the system must be in the state  given by the unitary evolution of the state $|B \r$. The probability interpretation is  only valid in  conjunction with a measurement.

 To see the problems that entail the extrapolation into unjustified realms of the standard interpretative scheme of quantum mechanics, let us consider in more detail the Schwinger process:  we start by noting that we must imagine the electric field to be turned on at some finite time, for otherwise, the problem of  the electron filed in interaction with an eternal electric filed, would be stationary, and if the system was prepared in its vacuum state, or  the state corresponding to the minimal eigenvalue of the full hamiltonian, it would remain in that state for all times and the issue of pair creation would not  make any sense .  So let us assume that the electric field is  turned on   at some time  $ t_1$ ( one is often interested in considering such process  to be  carried  out adiabatically , so we can imagine this taking place during a long interval $\Delta (t)$ centered at $t_1$  ). Analogously, we assume the electric field to be turned off at $t_2$. Let us assume that that electric field points in the x direction.  We now assume that the system is prepared at  a time $t<<t_1$ in the vacuum state ( $|in,0\r$) and we ask about the probability amplitude  for observing the system in the state containing an electron and a positron, in  the single  particle states $\psi_1$ and
   $\psi_2$ respectively.  This question has  a very well defined answer in quantum theory, which is  simply  $\l0; in | S |(\psi_1, -), (\psi_2,+);out \r  $ (of course this might not be easy to compute but that is quite a different issue).  However, let us note that we can not assume that the system, in the absence of a measurement,
   has  a well defined  probability of being in the state $|(\psi_1, -), (\psi_2,+);out \r  $, among other things, because the initial state is invariant under translations in the $y, z$ plane,  the hamiltonian preserves this  invariance but the state $|(\psi_1, -), (\psi_2,+);out \r  $ will not in general share such invariance. 
     Thus,   while  we are perfectly justified in viewing the S matrix calculation to yield the  prediction for probability for the observation of pair creation  out of the vacuum when contemplating the {\it measurement} of the number of pairs  at  a certain time, we are not justified  in regarding the state of a field as  being anything but  $U| B \r $ ( where U is the unitary evolution operator ) before a measurement is carried out. Similarly  in the early universe  inflationary context, we have no right to 
 view the state  of the universe as anything but $U| 0 \r $ in the  absence of a measurement. 

Again  some of the most conspicuous  and  clear contradictions   that arise when  we give ourselves the right to use  unjustified interpretational  extrapolations, can be seen  by considering the   
 standard B-EPR\cite{EPR} setup while entertaining the notion that the two particles might have a particular spin  orientation  in the absence (or before) a  measurement is carried out \cite{Mermin}.
 We could, for instance, decide to trace over  the spin D.O.F. of one of the  particles of the EPR pair and obtain, for the spin  D.O.F. of the other  particle, a diagonal density matrix, and be tempted to interpret this as indicating that the particle has one of the two spins orientations.  But we know this leads to contradictions.
  
I'll try to convince the reader  that this is not  necessarily ``just philosophy", and that the early universe offers the "Laboratory "  where  some  of the issues can,  at least in principle, be  studied. It is worth pointing out that, in fact,  the motivations for inflation itself are often criticized as being  "too philosophical".

\section{ What to do? Our approach.}

As we have seen, we need  a  paradigm that will allow us to consider  a transition from a H.\&I.  state  for the universe to a state that is neither isotropic nor homogeneous, but without relying on an external system to carry out a measurement and without reliance on our own limitations as observers to select  a set   of degrees of freedom to be considered as unobservable, and so forth. 
Our approach will follow in this case the suggestions by Roger Penrose  that quantum mechanical wave function collapse is an actual dynamical process where a system is forced to jump into  one of
 a certain collection of states, breaking,  in the process, the unitary evolution of quantum mechanics.  Moreover, these ideas suggest that  the fundamental theory of gravity\footnote{Here the suggestion is that such theory is not just the result of the  standard type of quantization as  applied to gravity, 
but  something involving a  radical change in our  paradigm of physical theories.} might play a central role in the actual physics of collapse, by forbidding certain types of superpositions of gravitational degrees of freedom to exist for more than a very brief time. These ideas are certainly a  bit too schematic to allow us to carry  a detailed analysis of the issues at  hand, so we will consider a concrete formalism inspired on them.   The idea  behind the scheme we will be using is that the  quantum gravity requires both the modification of our theory of gravitation and that  of quantum mechanics. In this scheme the  fundamental degrees  of
gravitation are not related to the metric degrees of freedom in  any simple  way, but instead the latter appear as effective degrees of freedom of a non-quantum effective theory. 
The degrees of freedom of other fields  whether  or not fundamental are to be considered as susceptible to a quantum treatment on their own,  an assumption that  would avoid problems when contrasting  this paradigm with the well established successes of quantum theory in  non-gravitational realms.

These ideas leave room to consider, therefore, situations where a quantum treatment of other fields would be appropriate but a classical  treatment of gravitation  would be justified.  That is the realm of semi-classical gravity that  we will assume to be valid for most of the time. However, this approximation would break down in association with the quantum mechanical jumps that are considered to be part of the underlying quantum theory containing gravitation.

In accordance with the ideas above we will use a semi-classical description of gravitation in interaction with quantum  fields as reflected  in the semi-classical Einstein's equation $ R_{\mu\nu} -(1/2) g_{\mu\nu} R =8\pi G  \l T_{\mu\nu}  \r $ whereas the other fields are treated in the standard quantum field  theory (in curved space-time) fashion. This is supposed to hold at all times except when  a quantum gravity induced collapse of the  wave function occurs, at which time,  the excitation of the  fundamental quantum gravitational degrees of freedom must be taken into account,  with the corresponding breakdown of the semi-classical approximation. The  possible breakdown  of the semi-classical approximation is  formally represented  by the presence of  a  term  $Q_{\mu\nu} $ in  the semi-classical Einstein's   equation which is supposed to become nonzero {\bf only}  during the  collapse of the quantum mechanical wave function of the matter fields. Thus we write
\be
\label{SemiCEQ}
  R_{\mu\nu} -(1/2) g_{\mu\nu} R +Q_{\mu\nu}  =8\pi G  \l T_{\mu\nu}  \r 
 \ee
 Thus, we consider the development of the state of the universe during the time at which the seeds of structure emerge to be initially  described by a H.\& I. state for the gravitational and matter D.O.F..  At some stage, the quantum state of the matter fields reaches a stage whereby the corresponding state for the gravitational D.O.F. is forbidden, and a quantum collapse of the matter field wave function is triggered.  This new state of the matter fields  does no  longer  need to share  the symmetries of the initial state, and
 by its connection to the  gravitational D.O.F. now  accurately described by Einstein's semi-classical  equation  leads to  a  geometry that is no longer homogeneous and isotropic.
 
\section{ The inflationary   origin of  cosmic structures:  the amended story, or the story after the Gospel}
\label{sec_main}
\smallskip
  The starting point of the  analysis  is as usual the action of a scalar field coupled to
gravity\footnote{We  will be using units where $c=1$ but will keep $\hbar$ (with units of Mass $ M$ times Length  $L$ ), and $G $ ( with units of $ L/M$ ) 
explicitly throughout the manuscript. The coordinates in the metric $\eta, x^i $ will have units of length $L$  but  the metric components, such as the scale 
factor $a$  will be dimensionless.  The field  $\phi$ has units of $(M/ L)^{1/2}$,  and the potential $V$ has  units of $M/L^3$}.
\be
\label{eq_action}
S=\int d^4x \sqrt{-g} \lbrack {1\over {16\pi G}} R[g] - 1/2\nabla_a\phi
\nabla_b\phi g^{ab} - V(\phi)\rbrack,
\ee
 where $\phi$ stands for the inflaton and $V$ for the 
inflaton's potential.
 One then splits both, metric and
scalar field into a spatially homogeneous (`background') part and an
in-homogeneous part (`fluctuation'), i.e. $g=g_0+\delta g$,
$\phi=\phi_0+\delta\phi$.

Th equations governing the background unperturbed Friedmann-Robertson universe  with line element
$ ds^2=a(\eta)^2\left[- d\eta^2 + \delta_{ij} dx^idx^j\right]$, and the homogeneous scalar field $\phi_0(\eta)$ are, the
scalar field equation, 
\begin{equation}
\ddot\phi_0 + 2 \frac{\dot a}{ a}\dot\phi_0 +
a^2\partial_{\phi}V[\phi] =0, \label{Scalar0}
\end{equation}
and  Friedmann's
equation
\begin{equation}
3\frac{\dot a^2}{a^2}=4\pi G  (\dot \phi^2_0+ 2 a^2 V[\phi_0]).
\end{equation}

 The background solution
 corresponds to the standard inflationary cosmology  which written using a conformal time,
 has, during the inflationary era,  a scale factor
$a(\eta)=-\frac{1}{H_{\rm I} \eta},$
 with $ H_I ^2\approx  (8\pi/3) G V$and with the scalar $\phi_0$ field in the slow roll regime so $\dot\phi_0= - ( a^3/3 \dot a)V'$. 
 This era is supposed to give rise to a reheating period whereby the universe is repopulated with ordinary matter fields, and then to a standard hot 
 big bang cosmology leading up to the present cosmological time. The functional  form  of $a(\eta)$ during these latter periods is,  of course, different but we
  will ignore such details on the account that  most of the change in the value of $a$ occurs during the inflationary regime.  The overall normalization of the scale 
  factor will be set so $ a=1$ at the "present  cosmological time". The inflationary regime would end for a value of $\eta=\eta_0$, negative and very small  in absolute terms.
 
 The perturbed metric can be written
\begin{equation}
ds^2=a(\eta)^2\left[-(1+ 2 \Psi) d\eta^2 + (1- 2
\Psi)\delta_{ij} dx^idx^j\right],
\end{equation}
 where $\Psi$  stands for the relevant perturbation and is called
the Newtonian potential.

The perturbation of the scalar field leads to a perturbation of the energy momentum tensor, and
thus Einstein's equations at lowest order lead to
\begin{equation}
\nabla^2 \Psi  =4\pi G \dot \phi_0  \delta\dot\phi= s   \delta\dot\phi,   
\label{main3}
\end{equation}
where we introduced the abbreviation $s=4\pi G \dot \phi_0$.

Now we consider the quantum theory of the field $\delta\phi$.
It is convenient  to  work with the  rescaled field variable  $y=a \delta \phi$ and its conjugate momentum $ \pi = \delta\dot\phi /a $. 
 To avoid  some  distracting  infrared problems  we set the problem in  a finite   box of side $L$, and  decompose the  real
field  and momentum operators  as
\begin{equation}
\y(\eta,\vec{x})=
 \frac{1}{L^{3}}\sum_{\vec k}\ e^{i\vec{k}\cdot\vec{x}} \hat y_k
(\eta), \qquad \py(\eta,\vec{x}) =
\frac{1}{L^{3}}\sum_{\vec k}\ e^{i\vec{k}\cdot\vec{x}} \hat \pi_k
(\eta),
\end{equation}
where the sum is over the wave vectors $\vec k$ satisfying $k_i L=
2\pi n_i$ for $i=1,2,3$ with $n_i$ integer and where $\hat y_k (\eta) \equiv y_k(\eta) \ann_k +\bar y_k(\eta)
\cre_{-k}$ and  $\hat \pi_k (\eta) \equiv g_k(\eta) \ann_k + \bar g_{k}(\eta)
\cre_{-k}$
with
\begin{equation}
y^{(\pm)}_k(\eta)=\frac{1}{\sqrt{2k}}\left(1\pm\frac{i}{\eta
k}\right)\exp(\pm i k\eta),\qquad
g^{\pm}_k(\eta)=\pm
i\sqrt{\frac{k}{2}}\exp(\pm i k\eta) . \label{Sol-g} 
\end{equation}

 As we will  be interested in considering a kind of self induced collapse which
 operates in close analogy with  a ``measurement", it is convenient to work
 with  Hermitian operators, which in ordinary quantum mechanics are the ones susceptible of direct measurement.
Thus we decompose both $\hat y_k (\eta)$ and $\hat \pi_k
(\eta)$ into their real and imaginary parts $\hat y_k (\eta)=\hat y_k{}^R
(\eta) +i \hat y_k{}^I (\eta)$ and $\hat \pi_k (\eta) =\hat \pi_k{}^R
(\eta) +i \hat \pi_k{}^I (\eta)$ where
\begin{equation}
\hat{y_k}{}^{R,I} (\eta) =
\frac{1}{\sqrt{2}}\left(
 y_k(\eta) \ann_k{}^{R,I}
 +\bar y_k(\eta) \cre{}^{R,I}_k\right) ,\qquad  
\hat \pi_k{}^{R,I} (\eta) =\frac{1}{\sqrt{2}}\left( g_k(\eta)
\ann_k{}^{R,I}
 + \bar g_{k}(\eta) \cre {}^{R,I}_{k} \right).
\end{equation}
We note that the operators $\hat y_k^{R, I} (\eta)$ and $\hat
\pi_k^{R, I} (\eta)$ are, therefore, hermitian operators.  
Note that the operators corresponding to $k$ and $-k$ are identical in the real
case (and identical, up to a sign, in the imaginary case).

We, now, proceed to calculate the commutator, and we will
find out that any of them are standard:
\begin{equation}
  [\hat{a}_k^R, \hat{a}_{k'}^R] =\hbar L^3\left( \delta_{k,k'} +
    \delta_{k,-k'}\right), \quad [\hat{a}_k^I, \hat{a}_{k'}^I] =
  \hbar L^3\left(\delta_{k,k'} - \delta_{k, -k'}\right)
\end{equation}

Next, we specify  the way we will model the  collapse, and follow the field evolution through  this collapse
to the end of inflation.
  We will assume that the collapse is
somehow analogous to an imprecise measurement\footnote{An imprecise measurement of an observable is one in which one does not end with an exact eigenstate  of that observable but  rather with a state which is  only peaked around the eigenvalue. Thus, we could consider measuring a  certain particle's position and momentum so as to end up with a state that is a wave packet with both position and momentum defined to a limited extent, and which,  of course, does not  entail a conflict with Heisenberg's uncertainty bound.} of the
operators $\hat y_k^{R, I}
(\eta)$ and $\hat \pi_k^{R, I} (\eta)$ which, as we pointed out are
hermitian operators and thus reasonable observables.  These field
operators contain complete information about
the field (we ignore here, for simplicity, the relations between the modes $k$ and $-k$).

 Let $|\Xi\rangle$ be any state in the Fock space of
$\hat{y}$. Let us introduce the following quantity:
$d_k^{R,I} = \l \ann_k^{R,I} \r_\Xi.$
Thus,  the expectation values of the modes are expressible
as
\begin{equation}
\l {\y_k{}^{R,I} (\eta)} \r_\Xi = \sqrt{2} \Re (y_k (\eta) d_k^{R,I}),  \qquad
\l {\py_k{}^{R,I} (\eta)} \r_\Xi = \sqrt{2} \Re (g_k (\eta) d_k^{R,I}).
\label{timeEv}
\end{equation}

For the vacuum state $|0\rangle$ we  have, of course,:
$
\l{\y_k{}^{R,I}}\r_0 = 0, \l\py_k{}^{R,I}\r_0 =0,
$
while their corresponding uncertainties are
\begin{equation}\label{momentito}
\fluc{\y_k {}^{R,I}}_0 =(1/2) |{y_k}|^2(\hbar L^3), \qquad
\fluc{\pf_k {}^{R,I}}_0 =(1/2)|{g_k}|^2(\hbar L^3).
\end{equation}


Now we will specify the rules according to which collapse happens.
Again, at this point our criteria will be simplicity and naturalness.

What we have to describe is the state $|\Theta\rangle$ after the
collapse.  It turns out that, for the goals at hand, all we need to specify 
$d^{R,I}_{k} = \langle\Theta|\ann_k^{R,I}|\Theta\rangle $.
In the vacuum state, $\y_k$ and
$\py_k$ individually are distributed according to Gaussian
distributions centered at 0 with spread $\fluc{\y_k}_0$ and
$\fluc{\py_k}_0$, respectively.  However, since they are mutually
non-commuting, their distributions are certainly not independent.  In
our collapse model, we do not want to distinguish one over the other,
so we will ignore the non-commutativity  and make the following
assumption about the (distribution of) state(s) $|\Theta\rangle$ after
collapse:
\begin{eqnarray}
\l {\y_k^{R,I}(\eta^c_k)} \r_\Theta&=&x^{R,I}_{k,1}
\sqrt{\fluc{\y^{R,I}_k}_0}=x^{R,I}_{k,1}|y_k(\eta^c_k)|\sqrt{\hbar L^3/2},\\
\l {\py_k{}^{R,I}(\eta^c_k)}\r_\Theta&=&x^{R,I}_{k,2}\sqrt{\fluc{\pyRI_k}
_0}=x^{R,I}_{k,2}|g_k(\eta^c_k)|\sqrt{\hbar L^3/2},
\end{eqnarray}
where $x_{k,1}^{R,I},x_{k,2}^{R,I}$ are  selected randomly from within a Gaussian
distribution centered at zero with spread one.
 Here, we must emphasize that our universe corresponds 
to a single realization of these random variables, and thus  each of these quantities 
has a  single specific value. From the equations above, we  solve for $d^{R,I}_k$, and 
using the result in Eq. (\ref{timeEv}) we obtain  $ {\y_k{}^{R,I} (\eta)} \r$ and
 $\l {\py_k{}^{R,I} (\eta)} \r $ for the   state  that resulted from the collapse for all later times.

 We should keep in mind that this specific recipe for the collapse is   just an example  among  the  simple  and natural ones, and that other possibilities do exist, and  those may lead to different
  predictions. In fact,  in \cite{Us}   an alternative  recipe was considered,  that seems to be quite  promising in dealing 
  with the fine tuning problem that  generically affect inflationary models \cite{Napflio}.
  
\section{Analysis of the Phenomenology}

Now, we must put together our semi-classical  description of  of the gravitational  D.O.F. and the quantum mechanics description of the inflaton field.  We recall that this entails the semi-classical version of 
the perturbed Einstein's equation that, in our case, reduces to:

\begin{equation}
\nabla^2 \Psi  = s \l  \delta\dot\phi \r = (s/a) \l \py  \r
\label{SemiEE}
\end{equation}

The Fourier components  at  the conformal time $ \eta$ are given by:
\begin{equation}
 \Psi_k ( \eta) = -(s/ak^2) \l {\py_k{}(\eta)}  \r
\label{SemiEEK}
\end{equation}
  Where the expectation value is the one corresponding to the appropriate state of the quantum field.  Thus, before the collapse, the state is the vacuum and, therefore,   $\Psi_k ( \eta)  =0$, and the space-time is  still homogeneous and isotropic at the corresponding scale, while after the collapse we have 
  \begin{equation} 
   \Psi_k ( \eta)  = -(s/ak^2) \l {\py_k{}(\eta)}\r_\Theta= -(s/ak^2) [ \l{\py_k{}^R(\eta)}\r_\Theta + i\l{\py_k{}^I(\eta)}\r_\Theta] 
    \label{SemiEEK2}
\end{equation}
    which no longer vanishes, indicating that after this time the universe became anisotropic and in-homogeneous at the corresponding scale.
 We now can reconstruct the space-time value of the Newtonian potential using
\begin{equation}
\Psi(\eta,\vec{x})=
 \frac{1}{L^{3}}\sum_{\vec k}\ e^{i\vec{k}\cdot\vec{x}} \Psi_k
(\eta),
\label{Psi}
 \end{equation} 
 which can be  used to extract the quantities of observational interest.
 The  measured quantity is essentially, through its imprint on the temperature fluctuations, the
``Newtonian potential" on the surface of last scattering: $
\Psi(\eta_D,\vec{x}_D)$.  This quantity is identified with the temperature fluctuations on the  surface of last scattering, by regarding those as  due to the gravitational red shift associated with the gravitational potential well from which the photons emerge\footnote{ The  gravitational change in 
frequency  $ \Delta(\nu)/\nu  = \Psi$,  leads through the black body radiation formula to $ \Delta(\nu)/\nu = \Delta (T)/T$ and therefore $\Delta (T)/T =\Psi$. }. From  this quantity, one 
extracts
\begin{equation}
\alpha_{lm}=\int \Psi(\eta_D,\vec{x}_D) Y_{lm}^* d^2\Omega.
\end{equation}
To evaluate the  expected value for the quantity of interest we use (\ref{SemiEEK2}) and (\ref{Psi}) to
write
\begin{equation}
 \Psi(\eta,\vec{x})=\sum_{\vec k}\frac{s  U(k)} {k^2}\sqrt{\frac{\hbar
k}{L^3}}\frac{1}{2a}
 F(\vec{k})e^{i\vec{k}\cdot\vec{x}},
\label{Psi2}
\end{equation}
where  the factor $U(k)$  represents the aspects of
the evolution of the quantity of interest associated with the
physical processes occurring during the period  from re-heating to decoupling, which include, among
  others, the acoustic oscillations of the plasma and which are not central to the issues we are exploring in this work. 

 Then, after some algebra one finds
\begin{eqnarray}
\alpha_{lm}&=&s\sqrt{\frac{\hbar}{L^3}}\frac{1}{2a} \sum_{\vec
k}\frac{U(k)\sqrt{k}}{k^2} F(\vec k)  4 \pi i^l  j_l((|\vec k|
R_D) Y_{lm}(\hat k),\label{alm1}
\end{eqnarray}
 where  $j_l(x)$ is the spherical Bessel function of the first kind, and where $\hat k$ indicates the direction of the vector $\vec  k$. It is in this
expression that the justification for the use of statistics becomes
clear.  The quantity we are in fact considering is the result of 
the combined contributions of an
ensemble of harmonic oscillators, each one contributing with a complex
number to the sum, leading to what is, in effect, a two dimensional random
walk whose total displacement corresponds to the observational
quantity. To proceed further, we must evaluate the most likely value
for such total displacement. This we do with the help of the imaginary
ensemble of universes, and the identification of the most likely value
with the ensemble mean value.  Note that this is used here only as a calculational tool to extract the most likely value of the "random walk". No other role is given to the ensemble and no call for the existence of 
its elements is invoked.  Now we
compute the expected magnitude of this quantity.  Taking the continuum limit, and  after some algebra \cite{ Us} we find, 
\begin{equation}
|\alpha_{lm}|^2_{M. L.} 
=\frac{s^2  \hbar}{2 \pi a^2} \int \frac {U(k)^2
C(k)}{k^4} j^2_l((|\vec k| R_D)  k^3dk, \label{alm4}
\end{equation}
where some of the information contained in $F(k)$ has become  encoded in:
\begin{equation}
C(k)\equiv 1+ (2/ z_k^2) \sin (\Delta_k)^2 + (1/z_k)\sin (2\Delta_k)
\label{ExpCk}
\end{equation}
where $\Delta_k= k \eta -z_k$, $ z_k =\eta_k^c k$ with $\eta$ represents the conformal  time of observation, and $\eta_k^c$ the conformal  time of collapse of the mode $k$.

  The last expression for $|\alpha_{lm}|^2_{M. L.}$ can be made more useful
by changing the variables of integration to $x =kR_D$ leading to
\begin{equation}
|\alpha_{lm}|^2_{M. L.}=\frac{s^2   \hbar}{2 \pi a^2} \int
\frac{U(x/R_D)^2 C(x/R_D)}{x^4}    j^2_l(x) x^3 dx,
\label{alm5}
\end{equation} This expression reveals that, if one ignores   late time physics processes represented in $U$ and the remaining signatures of the  collapse process represented in $C$, the observational spectrum should have no dependence on the size of the surface of last scattering $R_D$. 

 Turning our attention to the expression in Eq. (\ref{ExpCk}),we note  that the  appearance of  the time of observation could, in principle, lead to concerns, as that should be taken as the time of decoupling, which is thought to lie in a regime where inflation has long ended. In practice, its effects can be expected to be negligible because ( in our convention) that  conformal time would be  a an exponentially  small negative number ( i.e.  $\eta \to 0^{-} $ as $ a \to \infty$).
 Next  we note, in order to get a reasonable spectrum,
there seems to be only one simple option: that $z_k$ is essentially
independent of $k$,  that is, the time of collapse of the different
modes should depend on the mode's frequency according to
$\eta_k^c=z/k$. In fact,  the standard answer, in the absence of late time physical effects such as plasma oscillations,  would correspond to $C(k)=1$.
This is a rather strong conclusion which could  represent   relevant information about whatever the mechanism of collapse is, and  leads, as we will see,  to actual tests  on the  feasibility of different mechanisms for the physical collapse. 

  To end this section, we consider  a question raised  during the talk, by  W. Unruh. He  asked about the reason behind the fact that the  oscillatory behavior 
  reflected in $C(k)$ appears in the present analysis and not in the standard treatments, despite the fact that
   the system is linear and linear  averages can not be expected to hide any real feature.  To address this issue,
    going  beyond  the calculation itself requires a simple  analogy.  I believe that it is provided by the consideration  of a simple  harmonic oscillator initially  prepared on a highly excited eigen-state of the hamiltonian.
 If all our treatment focusses on carrying averages of quantities extracted directly from the initial state we will find no trace of any oscillatory behavior in time.  However, if we consider the state resulting of a partial or inaccurate  measurement of the position, which is considered to be  naturally described by a coherent state,
   the time of the measurement or collapse of the initial state into the coherent state will be  of relevance and from that point onwards oscillations in time can be expected for various of the relevant physical quantities.

\section{ A version of  `Penrose's  mechanism' for collapse in the cosmological  setting}
\label{sec_penrose}
\medskip
Penrose has advocated, for a long time,  that  the collapse of quantum
mechanical wave functions might be a dynamical process independent of observation, and that the
underlying mechanism might be related to gravitational
interaction. More precisely, according to this suggestion, collapse
into one of two quantum
mechanical alternatives would take place when the gravitational
interaction energy between the alternatives exceeds a certain
threshold. In fact, much of the initial motivation for the present
work came from Penrose's ideas and his questions regarding the quantum
history of the universe.

A very naive realization of Penrose's ideas in the present setting
could be obtained as follows: each mode would collapse by the
action of the gravitational interaction between its own possible
realizations. In our case, one could estimate the interaction energy
$E_I(k,\eta)$ by considering two representatives of the possible
collapsed states on opposite sides of the Gaussian associated with
the vacuum. We will denote the two alternatives by the indices $(1)$ and $(2)$. Clearly, we must  interpret $\Psi$  as the Newtonian
potential and, consequently, the right hand side of Eq.
(\ref{main3}), (after a rescaling by $a^{-2}$ to replace  the laplacian expressed in the  
comoving coordinates $x$ to
a laplacian associated with coordinates measuring physical length )  should be identified  with  matter density $\rho$.  Therefore, $\rho= a^{-2}\dot\phi_0 \dot \delta\phi  =a^{-3}\dot\phi_0\pi^y $. Then the  relevant  energy  is given by :
\be\label{GE1}
E_I(\eta)=\int \Psi^{(1)} \rho^{(2)}dV =\int \Psi^{(1)}(x,\eta) \rho^{(2)}(x,\eta)a^3 d^3 x = 
\int \Psi^{(1)}(x,\eta) \dot\phi_0 ( \pi^y (x,\eta) )^{(2)}d^3x. 
\ee 
 where $\Psi^{(1)} $ represents the Newtonian potential that would have arisen if the system had collapsed into the alternative $(1)$, and $ \rho^{(2)}$  represents the density perturbation associated with a collapse into the alternative $(2)$. Note that in this section we are ignoring the overall sign of this energy which being a  gravitational binding energy would  naturally be negative.
We next express this energy in terms of the Fourier expansion  leading to : 
\be
 E(\eta)= (1/L^6) \Sigma_{k, k'} \Psi_{k}^{(1)}( \eta)
\dot\phi_0 (\pi^y )^{(2)}_{k'} (\eta) \int  e^{i x (k-k')} d^3x = (1/L^3)\dot\phi_0 \Sigma_{k}
\Psi^{(1)}_{ k}( \eta) (\pi^y )^{(2)}_{k}(\eta) , 
\ee 
where $(1),(2)$
refer to the two different realizations chosen. Recalling
 that $\Psi_{ k} = ( s/ak^2) \pi^y_k$, with $s= 4\pi G\dot\phi_0$,  we find 
  \be
 E(\eta)= 4\pi G (a/L^3)
\dot\phi_0^2\Sigma_{k} (1/(ak)^2)
(\pi^y )^{(1)}_{k}(\eta)(\pi^y )^{(2)}_{k}(\eta), 
\ee 
  Using equation (\ref{momentito}), we  estimate $ (\pi^y_{k}) ^{(1)}(\eta) (\pi^y _{k})^{(2)}(\eta) $ by 
 $|<\pi^y_{k}> |^2 = \hbar k L^3 /4$,  and thus we obtain:
\be
 E_I(\eta) =  \Sigma_{k}( \pi \hbar  G/ak) (\dot\phi_0)^2.
 \ee
 which can be interpreted as the  sum of  the contributions of each mode to the interaction energy of different alternatives.
 According to all the considerations we have made,  we  view each mode's collapse as occurring independently, so the trigger  for the collapse of mode  $k$ would be, in  accordance to the  scheme based on  Penrose's ideas, the condition that  this energy  $
E_I(k,\eta)=( \pi \hbar  G/ak) (\dot\phi_0)^2 $ reaches the `one-graviton level',  namely, that  it  equals the value of the Planck Mass $M_p$.  Now we use the specific expressions for the scale factor  $ a=\frac{-1}{\eta H_I}$ and the slow rolling of the background scalar field $\dot \phi_0= -(1/3)  (a^3 /\dot a ) V'$ to  find 
\be \label{Emodek}
 E_I(k,\eta)=\frac{\pi \hbar  G}{ 9H_I^2} ( a/k) ( V')^2.
 \ee
Thus, the condition determining the time of collapse  $\eta^c_k$ of the mode $k$ is  that the above expression reaches the value $M_p$.  Thus, we find:
\be 
  z_k=\eta^c_k k =\frac{\pi }{9} (\hbar V')^2(H_I M_p)^{-3}=\frac{\epsilon} {8\sqrt {6\pi}}(\tilde V)^{1/2}\equiv  z^c,
 \ee
which is independent of $k$,  and, thus,  leads to a roughly  scale invariant spectrum of fluctuations in
accordance with observations.  It is worth pointing out that this result is far from trivial or expected on simple dimensional grounds, as the dimensionless factor $a$ could conceivable have appeared with the incorrect power in  Eq.( \ref{Emodek}). Moreover, we note that,  as the  energy of mode $k$ during the slow roll regime  is,  as shown  in Eq. (\ref{Emodek}), an increasing function of  conformal time $ \eta$, at the very early times the condition  for collapse would not  have been be fulfilled, and it is only as the universe expands that such point would be attained.

 One question that was raised during the presentation of this talk by T. Jacobson, concerns the finiteness  of energy created during each  collapse  and its independence of the "size" of the universe. In this regard, we note that the calculation of the energies above is analogous to the calculation of the vacuum energy of  a quantum field  in Minkowski space-time, in its ground state:  as each mode of the quantum field is essentially  an harmonic oscillator its  vacuum energy, regardless of the infinitude of the universe, is  a finite quantity : $(1/2) \hbar  \omega$. Similarly, in our case, the energy created in the collapse of each mode is finite. However, and in complete analogy with the standard quantum field considerations, the sum over all modes gives infinite  if the  number of modes is infinite.  This is in  both, the present case and the ordinary quantum field theory  case, an ultraviolet problem and has no relation with the universe's "size".

 Note that  the   formalism allows us to  look closer  into the collapse  issue and to ask for instance: when do the relevant modes collapse? 
    In order to answer  this question  we use the value for  $z^c$
and recall that the time of collapse
 is determined by $\eta^c_k =z^c/k $, and thus the scale factor at the time of collapse of the modes
with wave number $k$  was 
\be
     a^c_k= (H_I\eta^c_k)^{-1} = (12/\epsilon) k l_p (\tilde V )^{-1}
\ee 
     where $l_p$ stands for  the Planck length, and  $\epsilon \equiv (1/2) (M_{Pl}^2/ \hbar) (V'/V)^2 $ is the so called slow roll parameter of the inflationary model.
 As the  value of the scale factor $a$  at the last scattering surface was $a \approx 10^{-4}$ (recall that the scale factor $a$ has been set so its value today is $1$),  the modes that are relevant to say scales of order $10^{-3}$ of the size of the surface of last scattering (corresponding to a fraction of a degree in today's sky) have $k  \approx 10^{-10} {ly}^{-1}$. 
 
 Thus,  taking $\epsilon \times \tilde V $ to be of order $ 10 ^{-5}$, we have for  those modes     
 $
   a^c_k\approx 10^{-45}
$  
 corresponding to $ N_e =103$ e-folds of total expansion, or something like 80 e-folds before the end of inflation in  standard type of inflationary scenarios. Thus, in this scheme inflation must have at least 90 e-folds for it to include  the complete description of  the regime we are considering and to account also for the collapse of  the modes that are of the order of magnitude of  the surface of last scattering itself. The usual requirements of inflation put the lowest bound at something like 60 e-folds of inflation so the present requirement is not substantially stronger.
 
     This result can be directly compared   with the  so called, time of ``horizon crossing"  $ \eta^H_k $ for mode $k $, corresponding to the physical wavelength reaching the Hubble  distance $ H_I^{-1} $.
        Therefore this latter time is determined from: 
   \be
      a^H_k\equiv   a(\eta^H_k)= k/ (2\pi H_I) = k l_p (3/ 32 \pi^3)^{1/2} (\tilde V)^{-1/2}.
   \ee  
        Thus, the ratio of scale factors at collapse and at horizon crossing for a given mode is
$ a^c_k/ a^H_k= (16/\epsilon) (6 \pi^3)^{1/2}  (\tilde V)^{-1/2} $,
     which would ordinarily be a very  large number,  indicating that the collapse time would be much later than the time of  ``horizon exiting"  or crossing out,  of the corresponding mode.
     
Thus, we find that a naive realization of Penrose's ideas seems, at first sight, to be a good candidate to supply the element that we argued  
is missing in the standard accounts of the emergence of the seeds of cosmic structure from quantum fluctuations during the inflationary regime in the early universe.  However, more research along these lines is necessary to find out,  for instance, whether the scheme would imply a second collapse of modes already collapsed, and whether such secondary collapse could disrupt  in a substantial way the 
observational spectrum.

\section{Predictions and discussion}
   It is quite clear  
   that it would be very hard to find a scheme in which  the function $C(k)$ would be exactly  a constant, and that some  dependence of  $k$  will remain in any reasonable collapse scheme.   A particularly robust  source of this effect is  associated with the finite  time of decoupling as can be seen in
 in the expression for the function $C(k)$  \ref{ExpCk}. These dependences will, in turn, lead to slight deviations from the standard form of the spectrum before the inclusion of the late time physics such as  the plasma  oscillations.  This, in turn, can be expected to leave some traces in the  observational CMB  spectrum  that could, conceivable, be searched for observationally.
   
  However, the most striking  prediction of the scheme, is the absence of tensor modes, or  at least their  very strong suppression.   The  reason for this can be understood by considering the semiclassical version of Einstein's equation and its role in describing the manner in which the inhomogeneities and anisotropies in the metric arise  in our scheme.  As indicated in  the introduction, the metric is taken to be an effective description of the gravitational D.O.F., in the classical regime, and not as the fundamental D.O.F. susceptible to be  described at the  quantum level.  It is thus the matter degrees of freedom (which in the present context are represented by the inflaton field) the ones that 
    are described quantum mechanically  and which, as a result of a fundamental  aspect of gravitation at the quantum level,  undergo effective quantum collapse (the reader  should recall  that our point of view is that gravitation at the quantum level will be drastically different from standard quantum theories, and that, in particular, it will not involve universal unitary evolution). This leads to a nontrivial  value for $\l T_{\mu\nu} \r$, which leads to  the appearance of the metric fluctuations.  The point is that the 
energy momentum tensor  contains  linear and quadratic  terms in the expectation values of the 
quantum  matter  field fluctuations,  which are the source terms determining the  geometric perturbations. And a close examination  reveals that, in the case of the scalar perturbations, we have first order  contributions  proportional to $\dot\phi_0\l \delta\dot\phi  \r$ while no similar first order  terms appear as source of the tensor perturbations ( i.e. of the gravitational waves).  At  the time of the writing of this article, the tensor modes had not been detected, in contrast with the scalar modes, a fact that seems to lead support to the prediction above.    
 
 Next  we consider briefly the relationship between the present analysis and the  programs 
 that search for a quantum theory of gravity.   The first thing to  note is  that there is, in principle, no intrinsic conflict between the present analysis with any program that contemplates that the fundamental degrees of freedom are not necessarily directly tied with the observable space-time geometry. On the other hand, a conflict could arise if the quantum gravity theory required the geometric D.O.F. to be quantized on equal footing and in  all circumstances where  the matter degrees of freedom would require a quantum description, leading to a situation that would invalidate the applicability in the present, and similar  contexts,  of the semi-classical Einstein's equations.  The challenge for a theory of quantum gravity  that is posed by the present analysis is to  provide an   characterization of the circumstances in which that semiclassical  approach would be justified (which would include the  cosmological situation at hand ) while at the same time  provide  an explanation for the  effective mechanism of collapse that we  are attributing to some unknown aspect of quantum gravity.  

The search for manifestations of quantum aspects of gravitation, after being practically ignored 
for a long time, has transformed, as of late, into such an attractive enterprise that a large number of researchers have become  strongly attracted  even by  some ideas  of doubtful consistency and 
unclear interpretation. It seems  that in this  "Gold Rush", at least some very  interesting and direct avenues have been ignored. The  case of the emergence of the seeds of  cosmic structure, the only known process,  proved to be observationally accessible,  in which gravity and quantum physics seem to be inexorably tied,  seems to be the most  glaring example.
  On the other hand, it  is  naturally quite surprising, at least within the quantum gravity community,  that,  something that started as what could  be called  "purely philosophical considerations", would  lead to analysis that can be directly confronted with observations, and which give rise to predictions that could, in principle, invalidate aspects of the  emergent proposals.
  The lesson  we draw from this case  is that in the search for  clues of  aspects  of quantum gravity
 one has to face the most   obscure issues head on  rather than ignore them.  This is, of course true, in any scientific enterprise but it is more so  in a field such a quantum gravity, where there  are   
 so few pointers beyond  the  need  of  consistency as  one of the few reliable guiding principles.
 
 \section{Acknowledgments} I wish to thank the  conference organizers for the hospitality.
  This work was supported in part  by  the  grant  43914-F  of CONACyT .

\end{document}